\documentclass{article}

\pdfoutput=1
\usepackage{graphicx}
\usepackage[colorlinks=true, citecolor=blue, linkcolor=blue, urlcolor=blue]{hyperref}
\usepackage{amsmath, amsfonts, bbm}
\usepackage{color}
\usepackage{natbib}
\usepackage[english]{babel}
\usepackage{mathpazo}

\newtheorem{Proposition}{Proposition}[subsection]
\numberwithin{Proposition}{subsection}

\newcommand{\bs}{\boldsymbol}
\newcommand{\tr}{\operatorname{tr}}

\title{Objective Bayesian analysis for the multivariate skew-t model\footnote{\emph{Preprint submitted to Statistical Methods and Applications.}}}
\author{Antonio Parisi\\ \small{DEF, University of Rome Tor Vergata}\\ \small{\texttt{antonio.parisi@uniroma2.it}}
  \and
  Brunero Liseo\\ \small{MEMOTEF, Sapienza University of Rome}
}
\date{April 30, 2017}

\begin{document}

\maketitle

\begin{abstract}
We perform a Bayesian analysis of the $p$-variate \textit{skew-t} model, providing a new parameterization, a set of non-informative priors and a sampler specifically designed to explore the posterior density of the model parameters. Extensions, such as the multivariate regression model with skewed errors and the stochastic frontiers model, are easily accommodated. A novelty introduced in the paper is given by the extension of the bivariate \textit{skew-normal} model given in \cite{LisPar13} to a more realistic $p$-variate \textit{skew-t} model.
We also introduce the R package \texttt{mvst}, which allows to estimate the multivariate \textit{skew-t} model.\\

\end{abstract}

\section{Introduction}
In the last two decades there has been an explosion of interest around the possibility of constructing models which generalize the Gaussian distributions in terms of skewness and extra-kurtosis.
Interest can be partially explained with the empirical observations of phenomena, in different disciplines which could not be easily represented via Gaussian distributions. See \cite{gentbook} and \cite{acbook} for general accounts.
In this perspective, different proposals of skew-Student $t$ distributions have been proposed and now they play a prominent role as empirical models for heavy-tailed data, particularly in finance \citep{rachevbook}.

Among the various proposals we mention the \textit{skew-t} distribution obtained as a scale mixture of skew-normal densities \citep{ac03}; the ``two-piece'' $t$ distributions of \cite{han94} and \cite{FerSte99}; the \textit{skew-t} distribution arising from a conditioning argument (\citealp{dey:01}; \citealp{ac03}); the \textit{skew-t} distribution of \cite{jf03}, obtained by transforming a beta random variable, and the \textit{skew-t} distribution arising from a sinh-arcsinh transformation \citep{rosco}.
In practice, the most used of these are the Azzalini-type \textit{skew-t} distribution, in the form arising from scale mixing Azzalini's \textit{skew-normal} distribution \citep{ac03} and the ``{two-piece}'' $t$ distribution.

In the paper we will concentrate on the Azzalini-type \textit{skew-t} distribution. For a Bayesian analysis of the ``{two-piece}'' $t$ distribution one can refer to \cite{rubio} and \cite{leisen-skew} where a new objective prior is introduced for the degrees of freedom parameter.\\
Following \cite{acbook}, their version of the multivariate \textit{skew-t} distribution can be obtained as a scale mixture of multivariate \textit{skew-normal} distributions.
Let $W_0\sim SN_p(\bs{0},\bs{\Omega}, \bs{\alpha})$, where $\bs{\Omega} $ is the correlation matrix of the multivariate normal density appearing in the density of $W_0$, and $V\sim\Gamma(\nu/2, \nu/2)$.\\
Let $W=V^{-\frac 12}W_0$; integrating out $V$, one obtains the density of a $p$-variate \textit{skew-t} random vector as
\begin{equation}
\label{skew-dens}
 f_{\bs W}(\bs w; , \bs \alpha, \Omega, \nu) = 2\, t_p(\bs w; \nu) T_1\left(\bs\alpha' \bs w \left(\frac{\nu+p}{Q_{\bs w}+\nu}\right)^{1/2};\nu+p\right),
\end{equation}
where $Q_{\bs w}=\bs w^\prime \bs{\Omega}^{-1}\bs w$.

The joint estimation of the skewness vector $\alpha$ and the degrees of freedom parameter $\nu$ is hard even in the scalar case.
For the symmetric Student's $t$ distribution, it is known that the likelihood function tends to infinity
when $\nu$ goes to zero \citep{FerSte99}. \cite{fonseca} gave a condition for the existence of the MLE of $\nu$ in that
case. For the \textit{skew-t} distribution, the deviance approach has been implemented in \cite{ag:08}, where now the replacement of the MLE of $(\alpha, \nu)$ is based on the null hypothesis $H_0 : (\alpha, \nu) = (\alpha_0 , \nu_0 )$ 
and on a  $\chi_2^2$ distribution. However, simulation results have shown that this procedure provides only a partial solution to the problem. Alternatively, the modified score function approach has been applied to the \textit{skew-t} distribution by \cite{sartori}, although no proof of the finiteness of the resulting shape estimator has been provided; besides, this method requires the degrees of freedom parameter $\nu$ to be fixed. 
\cite{brgeli} provides an objective Bayesian solution to this problem in the scalar case.\\
In this paper we propose a method which generalizes both the results in \cite{brgeli} and \cite{LisPar13}. In fact we describe a Bayesian analysis of the $p$-variate \textit{skew-t} (\textit{ST}) model, providing a parameterization, a set of non-informative priors and a sampler specifically designed to explore the posterior density of the parameters of the model. Extensions of the model, such as the multivariate regression model with skewed errors and the stochastic frontiers model, are straightforward.\\
The main novelty of the present paper is given by the extension of the bivariate \textit{skew-normal} (\textit{SN}) model given in \cite{LisPar13} to a more realistic $p$-variate \textit{ST} model. Several issues arise in this extension, the most important of which is related to the elicitation of the prior distribution for the shape parameter and the sampling strategy for an additional set of latent variables.\\
This paper also introduces the R (\citealp{RCT15}) package {\texttt{mvst}}, which is available in the CRAN repository.\\ 
Several other packages are available for dealing with skew-symmetric distributions; among others, the R packages \texttt{sn} \citep{Azza15}, \texttt{EMMIXuskew} \citep{LeMc13}, \texttt{mixsmsn} \citep{PrBa13} and the Stata \citep{stata14} suite of commands \texttt{st0207} by \cite{MarGen2010}: however, most of them only rely upon the frequentist approach.

The rest of the paper is organized as follows: the second section introduces the model and the notation, along with the complete likelihood function and complete maximum likelihood estimators. It finally provides the prior distributions and the proof that the posterior distribution is proper.\\
The third section introduces the sampler and describes a set of proposal distributions.\\
Results from a simulation study are given in section four.

Throughout the paper, we will switch between three different parameterizations, characterized by the sets of parameters $\bs\theta_\star$, $\bs\theta^\star$ and $\bs\theta$; the former allows us to provide the proofs of our main results, the second one is the most sensible to elicit the prior distributions, while the latter is useful for the sampling strategy.

\section{The model}
The density of the multivariate \textit{skew-t} random vector has been given in \eqref{skew-dens}.
 For inferential purposes it is often necessary to introduce location and scale parameters, via the transformation $\bs Y=\bs \xi + \bs \omega \bs W$. We then finally say that a random vector $\bs Y$ is distributed as a $p$-variate \textit{skew-t} distribution, denoted by $\bs Y\sim ST_p(\bs\xi, \bs\alpha, \Sigma, \nu)$, if its pdf is given by
\begin{equation}\label{eq:alphapdf}
 f_{\bs Y}(\bs y; \bs\xi, \bs\alpha, \Sigma, \nu) = 2\, t_p(\bs y;\nu)\, T_1\left(\bs\alpha'\bs\omega^{-1}(\bs y-\bs\xi)\left(\frac{\nu+p}{Q_{\bs y}+\nu}\right)^{1/2};\nu+p\right),
\end{equation}
where $\bs\xi$ and $\bs \alpha$ are $p$-dimensional location and shape parameters, $\bs\omega$ is a diagonal matrix with the marginal scale parameters, so that $\Sigma = \bs\omega\Omega\bs\omega$ represents the scale matrix and $\nu$ represents the number of degrees of freedom. Moreover,
\begin{eqnarray*}
 Q_{\bs y} & = & (\bs y - \bs\xi)'\Sigma^{-1}(\bs y - \bs\xi),\\
 t_p(\bs y;\nu) & = & \frac{\Gamma((\nu+p)/2)}{|\Sigma|^{1/2}(\pi\nu)^{p/2}\Gamma(\nu/2)} 
 (1+Q_{\bs y}/\nu)^{-(\nu+p)/2}.
\end{eqnarray*}

There exist a useful stochastic representation of the random vector $\bs Y$ which is given in the following proposition.
\begin{Proposition}\label{prop:st}
Let
\begin{displaymath}
 \bs\delta = \frac{1}{(1+\bs\alpha'\Omega\bs\alpha)^{1/2}}\Omega\bs\alpha
\end{displaymath}
and let $I_A(\cdot)$ be the indicator function of the set $A$; define
\begin{equation}\label{eq:ZX}
\binom{Z}{\bs X} \sim N_{p+1}\left [\binom{0}{\bs 0}, \left(
 \begin{array}{cc}
  1 & \bs\delta^T\\
  \bs\delta & \Omega
 \end{array}\right)\right ],
\end{equation}
and
\begin{displaymath}
 \bs U = (-1)^{I_{(-\infty, 0)}(Z)} \bs X,
\qquad V\sim\Gamma(\nu/2, \nu/2),
\end{displaymath}
with $V$ independent of $U$. 
Then, (a) the random vector
\begin{displaymath}
 \bs Y = \bs\xi + \bs\omega \bs UV^{-1/2}\sim ST_p(\bs\xi, \bs\alpha, \Sigma, \nu)
\end{displaymath}
and (b) the joint density of $(\bs Y, Z, V)$ is given by
\begin{eqnarray}
\label{strep}
 f_{p+2}(\bs y, z, v) & = & f_p(\bs y\mid z, v) f(z) f(v) = N_p\left( \bs y \bs\xi + \bs\omega \bs\delta\frac{|z|}{\sqrt{v}}, \frac{1}{v} \bs\omega (\Omega - \bs\delta\bs\delta') \bs\omega\right)\\
  & \times & N_1(z, 0,1)\times \Gamma(v, \nu/2, \nu/2).\nonumber
\end{eqnarray}
\end{Proposition}
{\bf Proof:} the result is a direct consequence of the definition of the 
\textit{skew-t} distribution. 
Details can be found in Appendix \ref{proofa1}.

\subsection{Augmented likelihood function}
The above stochastic representation suggests to express the density of a \textit{skew-t} random vector as the marginal density of the augmented vector given in \eqref{strep}.\\
It is useful to define the parameter vectors $\bs\theta^\star = (\bs\xi, \bs\delta, \Sigma, \nu)$ and $\bs\theta = (\bs\xi, \bs\psi, G, \nu)$, where
\begin{eqnarray*}
 \bs\psi & = & \bs\omega\bs\delta,\\
 G & = & \bs\omega (\Omega - \bs\delta\bs\delta') \bs\omega = \Sigma - \bs\psi\bs\psi'.
\end{eqnarray*}
Using the new parameterization $\bs \theta$, and in the presence of a sample of $n$  i.i.d. observations $\bs{y}_i$ from a $p$-dimensional $ST(\bs\xi,\bs\psi,G,\nu)$, the augmented likelihood function is
\begin{eqnarray}\label{eq:augLike}
L(\bs\theta; \bs y, \bs z, \bs v) & \propto & \prod_{i=1}^n \left\{\phi_p\left(\bs y_i - \bs\xi - \bs\psi\frac{|z_i|}{\sqrt{v_i}}; \frac{1}{v_i}(\Sigma - \bs\psi\bs\psi')\right)\right.\nonumber\\
 & \times & \left.\phi_1(z_i;1)\times \Gamma\left(v_i;\frac{\nu}{2},\frac{\nu}{2}\right)\right\} =\\
  & = & \frac{\prod_{i=1}^n v_i^{p/2}}{|G|^{\frac n2}} \frac{(\nu/2)^{(n\nu/2)}}{(\Gamma(\nu/2))^n} \left(\prod_{i=1}^n v_i\right)^{\nu/2-1}\nonumber\\
  & \times & \exp\left\{-\nu/2\sum_{i=1}^n v_i\right\} \exp\left\{-\frac{1}{2}\sum_{i=1}^n z_i^2\right\}\nonumber\\
 & \times & \exp\left\{-\frac 12 \sum_{i=1}^n v_i \varepsilon_i' G^{-1} \varepsilon_i\right\},\nonumber
\end{eqnarray}
where $\bs z = (z_1, \dots, z_n)^\prime$, $\bs v = (v_1, \dots, v_n)^\prime$, $\varepsilon_i = \bs y_i-\bs\xi-\bs\psi\frac{|z_i|}{\sqrt{v_i}}$.

\subsubsection{Complete maximum likelihood estimators}\label{sec:CML}
The complete maximum likelihood (CML hereafter) estimators are obtained \textit{as if} we had observed the values of the latent variables $Z_i$'s and $V_i$'s. We will make use of the CML estimatates for the initialization of the sampling strategy, described below. They incorporate an additional piece of information, hence they could also be useful as a benchmark to evaluate and compare different estimators in a simulation experiment.\\

Given $\bs z$ and $\bs v$, the likelihood \eqref{eq:augLike} gets transformed into 
\begin{eqnarray*}
 L(\bs\theta ; \bs y, \bs z, \bs v)  & \propto & |G|^{-n/2} \exp\left\{-\displaystyle\frac 1 2\sum_{i=1}^n v_i\, \varepsilon_i'\, G^{-1}\, \varepsilon_i\right\}\\
  & \times & \displaystyle\frac{(\nu/2)^{(n\nu/2)}}{(\Gamma(\nu/2))^n} \left(\prod_{i=1}^n v_i\right)^{\nu/2-1} \exp\left\{-\nu/2\sum_{i=1}^n v_i\right\}
\end{eqnarray*}
After straightforward calculations, the CML estimators are obtained as:
\begin{eqnarray*}
 \hat{\bs\psi}_{CML} & = & \displaystyle\frac{1}{(\sum_{i=1}^n z_i^2)(\sum_{i=1}^n v_i) - (\sum_{i=1}^n |z_i|\sqrt{v_i})^2}\\
 & \times & \left[\left(\sum_{i=1}^n v_i\right)\left(\sum_{i=1}^n |z_i|\sqrt{v_i} \bs y_i\right) - \left(\sum_{i=1}^n |z_i|\sqrt{v_i}\right)\left(\sum_{i=1}^n v_i \bs{y}_i\right)\right],\\
 \\
  \hat{\bs\xi}_{CML} & = & \frac{1}{\sum_{i=1}^n v_i}\left[\left(\sum_{i=1}^n v_i \bs{y}_i\right) - \hat{\bs\psi}_{CML} \left(\sum_{i=1}^n |z_i|\sqrt{v_i}\right)\right],\\
 \\
 \hat G_{CML} & = & \frac 1 n \sum_{i=1}^n v_i\, \hat\varepsilon_i\, \hat\varepsilon_i',
\end{eqnarray*}
where
\begin{displaymath}
 \hat\varepsilon_i = \bs y_i-\hat{\bs\xi}_{CML}-\hat{\bs\psi}_{CML}\frac{|z_i|}{\sqrt{v_i}}.
\end{displaymath}
The estimator for $\nu$ have not a closed form expression: it is the solution of the following equation
\begin{displaymath}
 n\log(\hat\nu_{CML}/2) - n\, \psi(\hat\nu_{CML}/2) = \sum_{i=1}^n v_i - \sum_{i=1}^n\log(v_i) - n,
\end{displaymath}
where $\psi(\cdot)$ denotes the digamma function.

\subsection{Prior distributions}
\label{priors}
We assume the following prior structure for the parameters
\begin{displaymath}
 \pi(\bs\theta^\star) = \pi(\bs\xi) \pi(\bs\delta, \Sigma) \pi(\nu).
\end{displaymath}
As pointed out in \cite{LisPar13}, when $p>1$,  even following an objective Bayesian approach, $\bs\delta$ and $\Sigma$ cannot be considered a priori independent of each other. This depends on the expression of $G=\bs \omega (\Omega - \bs \delta \bs \delta^\prime) \bs \omega$: in order to guarantee the positive definiteness of $G$, one should consider, both in the analytical expression and in the computations, the constraint $\Omega - \bs \delta \bs \delta^\prime \succ 0$.\\
We further consider the decomposition
\begin{displaymath}
 \pi(\bs\delta, \Sigma) = \pi(\bs\delta|\Sigma) \pi(\Sigma)
\end{displaymath}
and we assume a flat prior for $\bs\xi$ and a conjugate Inverse Wishart prior for $\Sigma$. This way we adopt  the ``usual'' objective priors for the location and scale parameters as in the multivariate Normal model, which is nested in the multivariate \textit{ST} model, as $\bs \delta= \bs 0$ and $1/\nu \to 0$. In practice, we set 
\begin{displaymath}
 \begin{array}{l}
   \pi(\bs\xi) \propto 1\\
   \Sigma \sim IW(m , \Lambda )
 \end{array}
\end{displaymath}
In real applications, we will take $m=0$ and $\Lambda = \bs 0$. In \S\ref{sec:properPost}, we prove that the use of an improper prior on $(\bs\xi, \Sigma)$ produces proper posterior distributions, provided that the prior on the degrees of freedom parameter $\nu$ is proper and discrete over $\mathbb{N}$. Then we assume a uniform prior for $\nu$ over a set of 20 values ranging from 1 to 100.\\
Finally, we need to specify $\pi(\bs \delta \vert \Sigma)$. For each value of $\Sigma$, the parameter $\bs\delta$ lies in a $p$-dimensional region whose shape only depends on $\Sigma$ or 
$\Omega$. In particular, given the expression of $\delta$, it is easy to verify that 
\begin{equation}
 \bs\delta'\Omega^{-1}\bs\delta < 1,
 \label{eq:constraint}
\end{equation}
must hold, so the conditional parameter space is an ellipsoid, say $\Delta_\Sigma$, given by expression \eqref{eq:constraint}, centered at the origin and contained in the hyper-cube $(-1,1)^p$.
In any simulation based approach care must be taken that the proposed values actually satisfy \eqref{eq:constraint}. For computational convenience we prefer to directly include this constraint on the prior.
In the bivariate case, \cite{LisPar13} used an approximation of the Jeffreys' prior, normalized over $\Delta_\Sigma$. 
This normalization step, for large $p$, may become computationally demanding. For this reason, we propose to adopt a uniform prior over $\Delta_\Sigma$, whose volume can be evaluated in a closed form, so the normalizing constant is analytically tractable. Then we assume:
$(\bs\delta|\Sigma) \sim U(\Delta_{\Sigma}),$
that is
\begin{displaymath}
 \pi(\bs\delta|\Sigma) = \left(\frac{\pi^{p/2}}{\Gamma(p/2 + 1)}\sqrt{|\Omega|}\right)^{-1}.
\end{displaymath}
In the practical application of the \textit{ST} model,  we will use the $\bs\theta$ parameterization for our sampling strategy. Hence, we need to compute the Jacobian of the transformation $\bs\theta^\star\rightarrow \bs \theta$, which is given by
\begin{displaymath}
 |J| = \prod_{j=1}^p (G_{jj} + \psi_j^2)^{-1/2}.
\end{displaymath}

\subsection{Posterior propriety}\label{sec:properPost}
\begin{Proposition}
The posterior distribution of the model is proper.
\end{Proposition}
{\bf Proof:} Let $\bs\theta_\star = (\bs\xi, \bs\alpha, \Sigma, \nu)$, using the parameterization in \eqref{eq:alphapdf},
\begin{eqnarray*}
 \pi(\bs\theta_\star|\bs y) & = & \pi(\bs\theta_\star) \prod_{i=1}^n \left[2\, t_p(\bs y_i;\nu)\phantom{\frac{1}{1}}\right.\\
  & \times & \left. T_1\left(\bs\alpha'\bs\omega^{-1}(\bs y_i-\bs\xi)\left(\frac{\nu+p}{Q_{y_i}+\nu}\right)^{1/2}; \nu+p\right)\right]. 
\end{eqnarray*}
Since the c.d.f. $T_1(\cdot)$ is bounded by 1, one obtains
\begin{displaymath}
 \pi(\bs\theta_\star|\bs y) \leq\bar\pi(\bs\theta_\star|\bs y) = \pi(\bs\xi)\pi(\Sigma)\pi(\bs\alpha|\Sigma)\pi(\nu) \prod_{i=1}^n \left[2\, t_p(\bs y_i;\nu)\right].
\end{displaymath}
Notice that the parameter $\bs\alpha$ only appears in the prior distribution; then it can be integrated out to obtain
\begin{displaymath}
 \bar\pi(\bs\xi, \Sigma, \nu|\bs y) = \pi(\bs\xi)\pi(\Sigma)\pi(\nu) \prod_{i=1}^n \left[2\, t_p(\bs y_i;\nu)\right].
\end{displaymath}
The above expression is proportional to the posterior density of the parameters of a multivariate Student-$t$ model, with priors given as in \S\ref{priors}.
Theorem 1 in \cite{FerSte99} then guarantees that the posterior distribution of our \textit{ST} model is proper as soon as the prior on $\nu$ is proper and  $n\geq p+1$ except, possibly, for a set of Lebesgue measure zero in $\mathbb R^{n\times p}$. The finite precision of the data recording process can lead, under some choices for the prior distributions, to improper posterior distributions. However, it is possible to verify this condition for any given dataset, and we refer to the cited article for details.

\section{The sampler}
In the following, we describe the sampling strategy. We have used a Population Monte Carlo algorithm (PMC hereafter, see \citealp{MR2109057}), which improves and generalizes the one used in \cite{LisPar13} for the bivariate \textit{SN} model.\\
As a Monte Carlo method, the PMC sampler doesn't rely on convergence arguments, hence it can overcome the problem of multimodality of the posterior distribution; moreover, it offers a great flexibility in choosing the proposal density functions. For example, we use (approximations of) the full conditional distributions as proposal densities.\\

The outline of the algorithm for the \textit{ST} model is as follows:\\
\rule{\textwidth}{1pt}
\begin{itemize}
 \item At iteration 0, a population of $N$ particles $\bs\eta^{(0)}_{1:N}$, containing the values of $\bs\theta^{(0)}_{1:N}$, $\bs z^{(0)}_{1:N}$ and $\bs v^{(0)}_{1:N}$, is initialized. A possible initialization is described in \S\ref{inits}.
 \item At a generic iteration $t$
 \begin{itemize}
  \item new values for the particles are proposed following a proposal distribution $q(\bs\eta^{(t)})$, whose parameters possibly depend on the populations of particles in the previous iterations,
  \item the importance weights are computed as
  \begin{eqnarray*}
   \tilde\zeta_j^{(t)} & = & \tilde\pi(\bs\eta_j^{(t)}|\bs y) / q(\bs\eta_j^{(t)})\\
   \zeta_j^{(t)} & = & \tilde \zeta_j^{(t)} / \sum_{j=1}^N \tilde\zeta_j^{(t)}
  \end{eqnarray*}
  where $\tilde\pi$ and $\tilde\zeta$ denote the unnormalized posterior density function and importance weights.
  \item A set of quantities are obtained on the basis of the current particles and weights. This set includes the estimates of the parameters $\bs{\eta}^{(t)}$, a quantity related to the performance of the sampler in the $t$-th iteration
  \begin{displaymath}
   H^{(t)} = -\sum_{j=1}^N \zeta_j^{(t)}\log(\zeta_j^{(t)}),
  \end{displaymath}
  and all the other objects of interest.
  \item the particles $\bs\eta^{(t)}_{1:N}$ are multinomially resampled using the weigths $\bs\zeta^{(t)}$.
 \end{itemize}
 \item After $T$ iterations, the final estimates are obtained as a weighted mean of the estimates $\tilde{\bs\eta}^{(1:T)}$ with (unnormalized) weights given by $H^{(1:T)}$.
\end{itemize}
\rule{\textwidth}{1pt}
A quantity of special interest which can be easily obtained using the PMC is the marginal likelihood of each model. It can be estimated as
\begin{equation}
 \hat p(\bs y)\approx \frac{\sum_{t=1}^T H^{(t)}\sum_{j=1}^N \tilde\zeta_j^{(t)}}{N\sum_{t=1}^T H^{(t)}}.
 \label{eq:py}
\end{equation}

\subsection{Initial values for parameters}\label{inits}
The initial points are sampled by mimicking the stochastic representation of the model. Then
\begin{enumerate}
 \item the values of $\nu^{(0)}_{1:N}$ are sampled from the prior distribution;
 \item given $\nu^{(0)}_{1:N}$ the values of the latent variables $\bs z^{(0)}_{1:N}$ and $\bs v^{(0)}_{1:N}$ are sampled by the respective sampling distributions described in Proposition \ref{prop:st};
 \item given $\nu^{(0)}$, $\bs z^{(0)}_{1:N}$ and $\bs v^{(0)}_{1:N}$, the parameters $\bs\xi^{(0)}_{1:N}$, $\bs\psi^{(0)}_{1:N}$ and $G^{(0)}_{1:N}$ are obtained as the CML estimates of the parameters, as described in \S\ref{sec:CML}.
\end{enumerate}

\subsection{Proposals}
For the common parameters of \textit{SN} and \textit{ST} models, the proposal distributions are similar to those reported in \cite{LisPar13}; our versions 
are given in appendix \ref{Appx:propDist}. The \textit{ST} model, however, also includes the parameter $\nu$ and the latent variables $V_i$'s.\\
The parameter $\nu$ assumes values on a finite set, hence it is easy to simulate from its full conditional distribution
\begin{displaymath}
 \pi(\nu|\cdots) \propto \frac{(\nu/2)^{n\nu/2}}{(\Gamma(\nu/2))^{n}} \left(\prod_{i=1}^n v_i\right)^{\frac{\nu}{2}-1} \exp\left\{-\frac{\sum_{i=1}^n v_i}{2}\nu\right\}.
\end{displaymath}
Instead, to our knowledge, there is no simple way to draw values from the full conditional distribution of $V_i$, which is given by
\begin{displaymath}
 \pi(v_i|\cdots) = \frac{1}{k_{v_i}} v_i^{C-1} \exp\left\{-A_i v_i - B_i\sqrt{v_i}\right\},\qquad v_i > 0
\end{displaymath}
where
\begin{eqnarray*}
 A_i & = & 0.5[\nu + (\bs y_i-\bs\xi)' G^{-1} (\bs y_i-\bs\xi)]\\
 B_i & = & -(\bs y_i-\bs\xi)' G^{-1} \bs\psi|z_i|\\
 C & = & (\nu+p)/2
\end{eqnarray*}
and $k_{v_i}$ is the normalizing constant.\\
When $B_i = 0$ (for example in the symmetric case, where $\bs\psi$ is a null vector), then the full conditional for $v_i$ has a Gamma distribution. Otherwise, the sign of $B_i$ determines the right tail behaviour: when $B_i$ is positive (negative), the right tail of the full conditional distribution is thicker (lighter) than the right tail of a Gamma distribution.\\
Hence, we cannot propose values from a Gamma distribution, as it could jeopardize the validity of the method when $B_i<0$. On the other hand, proposing from a distribution with a thick tail could represent a huge loss in the efficiency of the sampler. For these reasons, we propose values using a rejection sampler (see, for example, \citealp{RoCa04}, \S 2.3) having the full conditional distribution as target density. We will
\begin{enumerate}
 \item define the distribution of the instrumental variable of the rejection sampler;
 \item choose the parameters of this distribution by minimizing the Kullback-Leibler divergence with respect to the target distribution;
 \item obtain the constant $M$ required by the rejection sampling algorithm;
 \item obtain the normalizing constant $k_{v_i}$, required by the PMC algorithm.
\end{enumerate}
Details are as follows:
\begin{enumerate}
 \item define $W = R^2$, with $R\sim \Gamma(\alpha_v, \beta_v)$; the instrumental density function is
 \begin{displaymath}
  f(w|\alpha_v, \beta_v) = \frac{\beta_v^{\alpha_v}}{2\Gamma(\alpha_v)} w^{\alpha_v/2-1} \exp(-\beta_v\sqrt{w});
 \end{displaymath}
 this density has a right tail which is thicker than the one of the target distribution;
 \item If we set $\alpha^\star_v = 2 C$ (see Appendix~\ref{Appx:RS}), the $KL(f||\pi_{v_i})$ divergence, as a function of $\beta_v$, has a minimum (in $\mathbb R^+$) in
 \begin{displaymath}
  \beta_v^\star = \frac{1}{2} \left(B_i + \sqrt{B_i^2 + 8 A_i (2C+1)}\right).
 \end{displaymath}
 Using the parameters $\alpha_v^\star$ and $\beta_v^\star$ we will optimise the efficiency of the rejection sampler.
 \item the Rejection Sampling algorithm requires a constant $M$ for which
 \begin{displaymath}
  \pi(v_i|\cdots)\leq M f(v_i).
 \end{displaymath}
 The value of $M$ can be found by defining the ratio $m(v_i) = \pi(v_i|\cdots) / f(v_i)$; given the parameters of the instrumental density, this function has a maximum in
 \begin{displaymath}
  v_i^\star = \left(\frac{\beta_v^\star-B_i}{2A_i}\right)^2.
 \end{displaymath}
 The value of $M$ can be finally obtained as $m(v_i^\star)$.
 \item To obtain the value of
 \begin{displaymath}
  k_{v_i} = \int_{\mathbb{R}^{+}} v_i^{C-1} \exp\left\{-A_i v_i - B_i\sqrt{v_i}\right\}dv_i,
 \end{displaymath}
 we use eq. 3.462 (1) in \citeauthor{GrRy94} (\citeyear{GrRy94}, GR hereafter), with $\nu=2C>0$, $\beta=A_i>0$, $\gamma=B_i$,
 \begin{displaymath}
  k_v = 2 (2A_i)^{-C}\Gamma(2C)\exp\left\{-\frac{B_i^2}{8A_i}\right\}D_{-2C}\left(\frac{B_i}{\sqrt{2A_i}}\right)
 \end{displaymath}
 where $D_p(z)$ is the parabolic cylinder function (GR, eq. 9.240) with $p=-2C$ and $z=B_i/\sqrt{2A_i}$, hence
 \begin{eqnarray}
  D_{-2C}\left(\frac{B_i}{\sqrt{2A_i}}\right) & = & \left[\frac{\sqrt{\pi}}{\Gamma\left(\frac{1+2C}{2}\right)} \Upsilon\left(C,\frac 1 2;\frac{B_i^2}{4A_i}\right) - \frac{\sqrt{2\pi}\frac{B_i}{\sqrt{2A_i}}}{\Gamma(C)}\Upsilon\left(\frac{1+2C}{2},\frac 3 2;\frac{B_i^2}{4A_i}\right)\right]\nonumber\\
   & \times & 2^{-C}\exp\left\{\frac{B_i^2}{8A_i}\right\}
  \label{eq:parabolic}
 \end{eqnarray}
 where $\Upsilon(\alpha,\gamma;z)$ denotes the confluent hypergeometric function (GR, eq. 9.210).
\end{enumerate}

\section{Simulation study}
\begin{figure}[ht]
 \centering
 \includegraphics[width=0.7\textwidth]{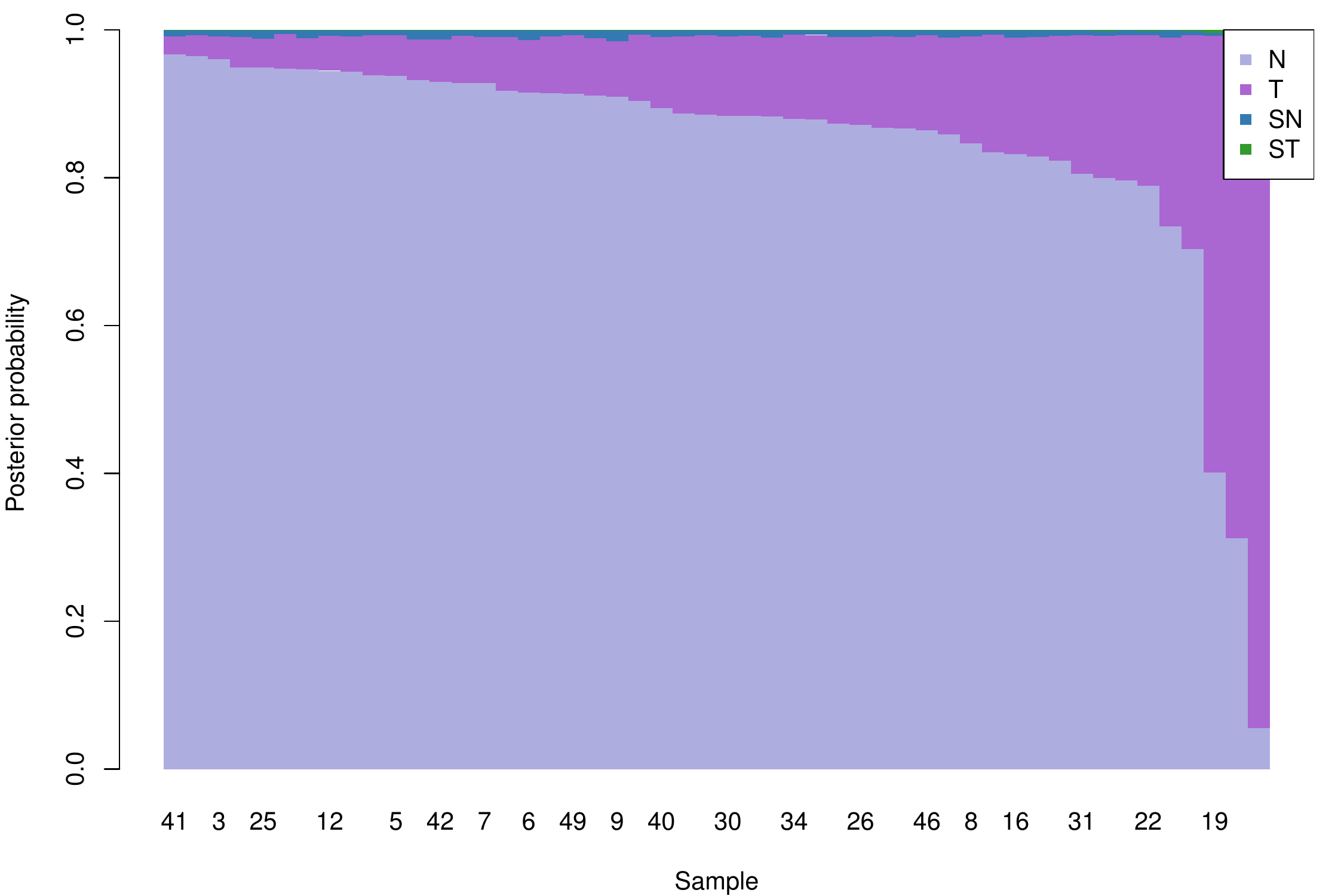}
 \caption{Simulation results for the Normally distributed samples.}
 \label{fig:postProbsN}
\end{figure}
In this section we use simulated data to evaluate the performances of the proposed approach. Since the multivariate \textit{ST} model may be considered an encompassing model including, as special cases, the multivariate Student-$t$ model, the multivariate \textit{SN} model and the multivariate normal one, it is of primary importance to verify the ability of the proposed approach to discriminate among these nested models.\\
For each of the four models, we have generated 50 samples; for each sample, we compute the posterior probabilities of each candidate model. These posterior probabilities are estimated using \eqref{eq:py} together with a uniform prior over the model space. 

In our simulations, each sample consists of $n=300$ observations with $p=4$ and
\begin{displaymath}
 \bs\xi = (5, 9, 3, 10)',\qquad\Sigma = \left(
 \begin{array}{rrrr}
  7 & 2 & 1 & 1\\
  2 & 8 & -2 & 3\\
  1 & -2 & 5 & -2\\
  1 & 3 & -2 & 8
 \end{array}\right).
\end{displaymath}
Samples from the \textit{SN} and \textit{ST} models have been generated using $\bs\alpha = (4, 4, 4, 4)'$. Data generated from the Student-$t$ and \textit{ST} models have $\nu=10$.

For each sample we have run the PMC algorithm using 20000 particles for each of 6 iterations. Results are summarized in the following four plots.
\begin{figure}[ht]
 \centering
 \includegraphics[width=0.7\textwidth]{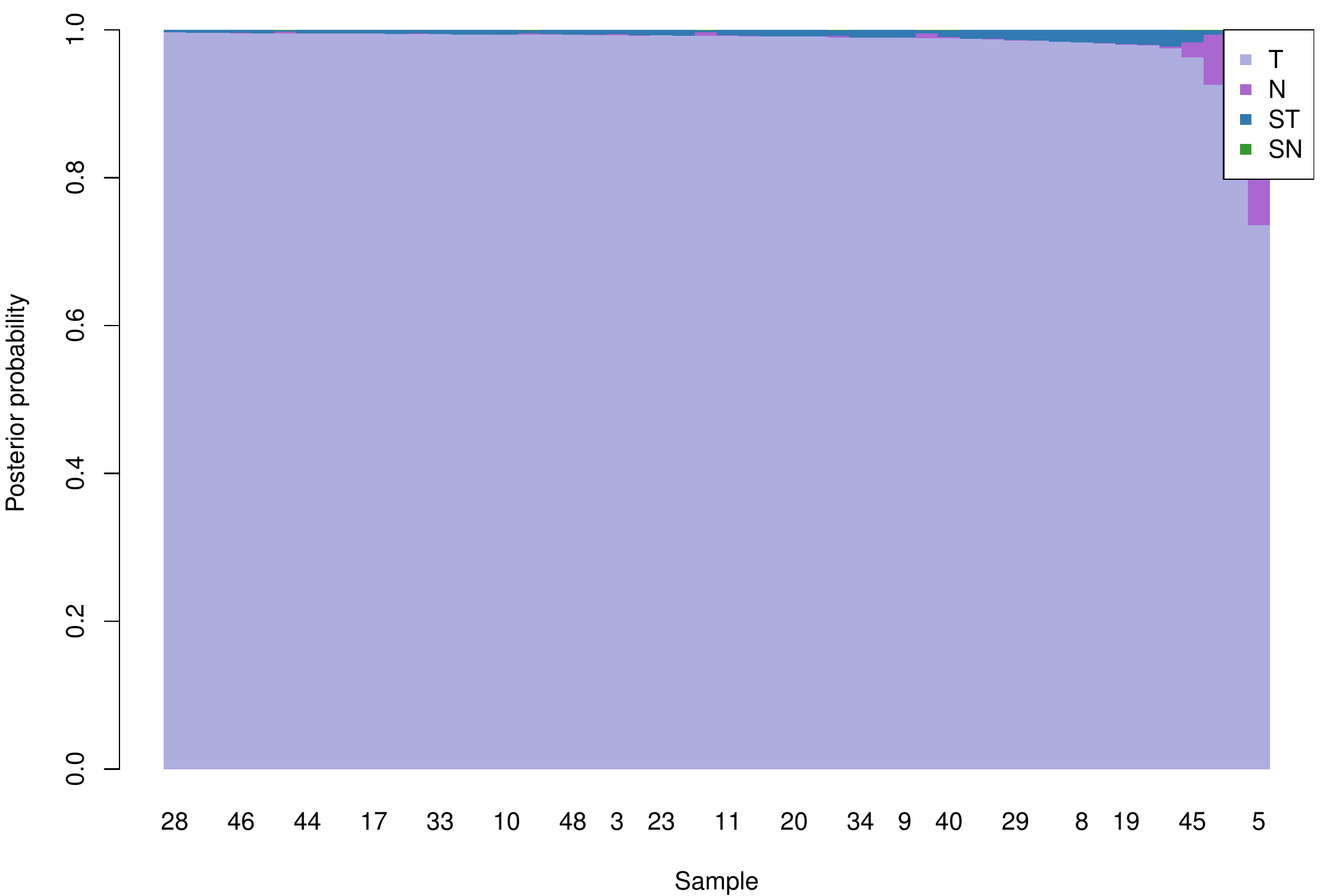}
 \caption{Simulation results for the Student-$t$ distributed samples.}
 \label{fig:postProbsT}
\end{figure}
\newline The barplot in Fig.\,\ref{fig:postProbsN} depicts the results for the Normally distributed samples. Each column stacks the posterior probabilities of the 4 candidate models estimated for a single sample. To improve the readability of the plot, bars have been rearranged in order to have a decreasing probability for the true model. Here, the true model is correctly identified in 47 cases; in the remaining cases (3 our of 50), the Student-$t$ model is preferred. The posterior probabilities for the remaining models are always very low.\\
\begin{figure}[ht]
 \centering
 \includegraphics[width=0.7\textwidth]{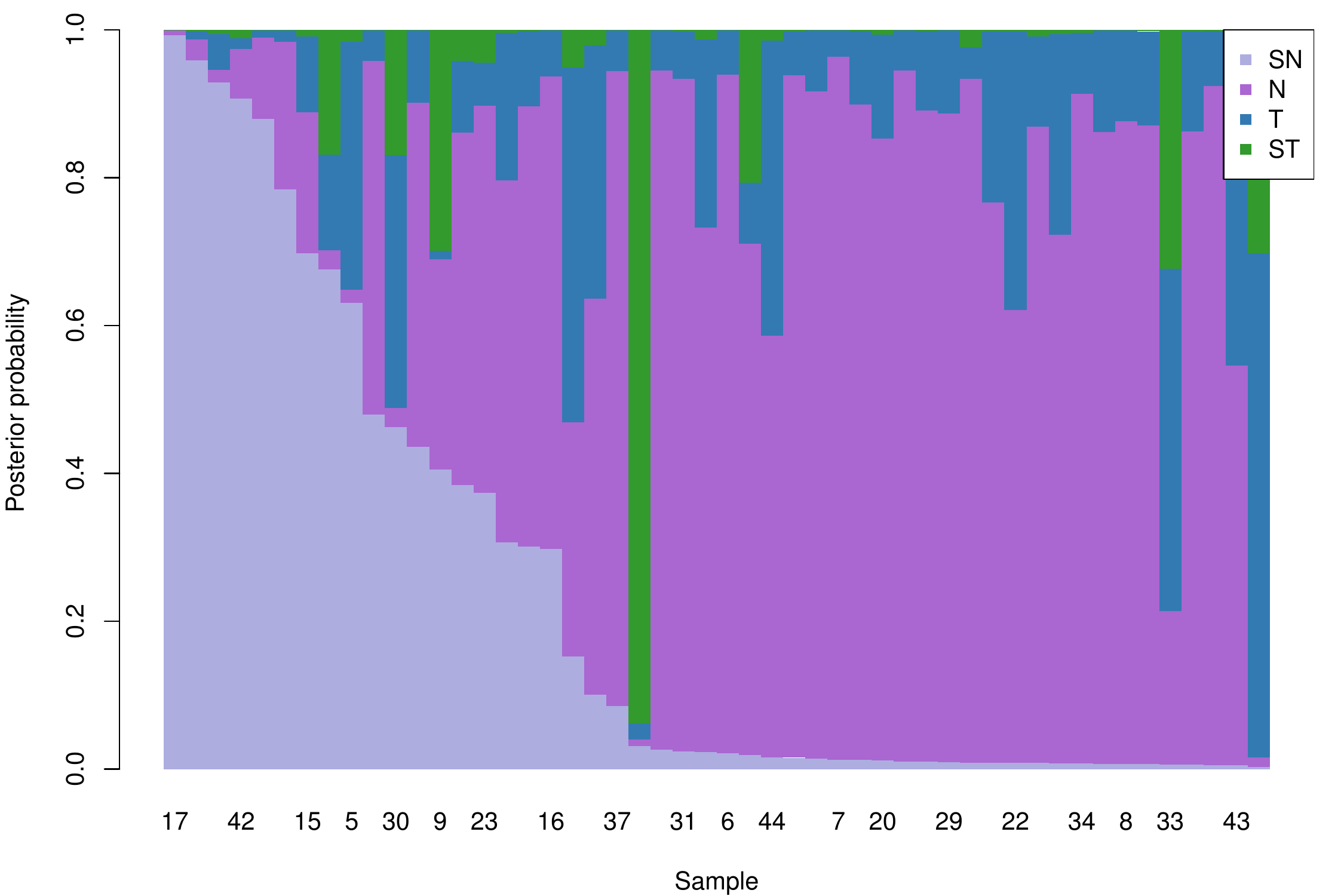}
 \caption{Simulation results for the \textit{SN}-distributed samples.}
 \label{fig:postProbsSN}
\end{figure}
\newline The situation is even more extreme when data come from a Student-$t$ distribution (Fig.\,\ref{fig:postProbsT}): here the true model is always correctly identified, with small to negligible probabilities for the other models.\\
\begin{figure}[ht]
 \centering
 \includegraphics[width=0.7\textwidth]{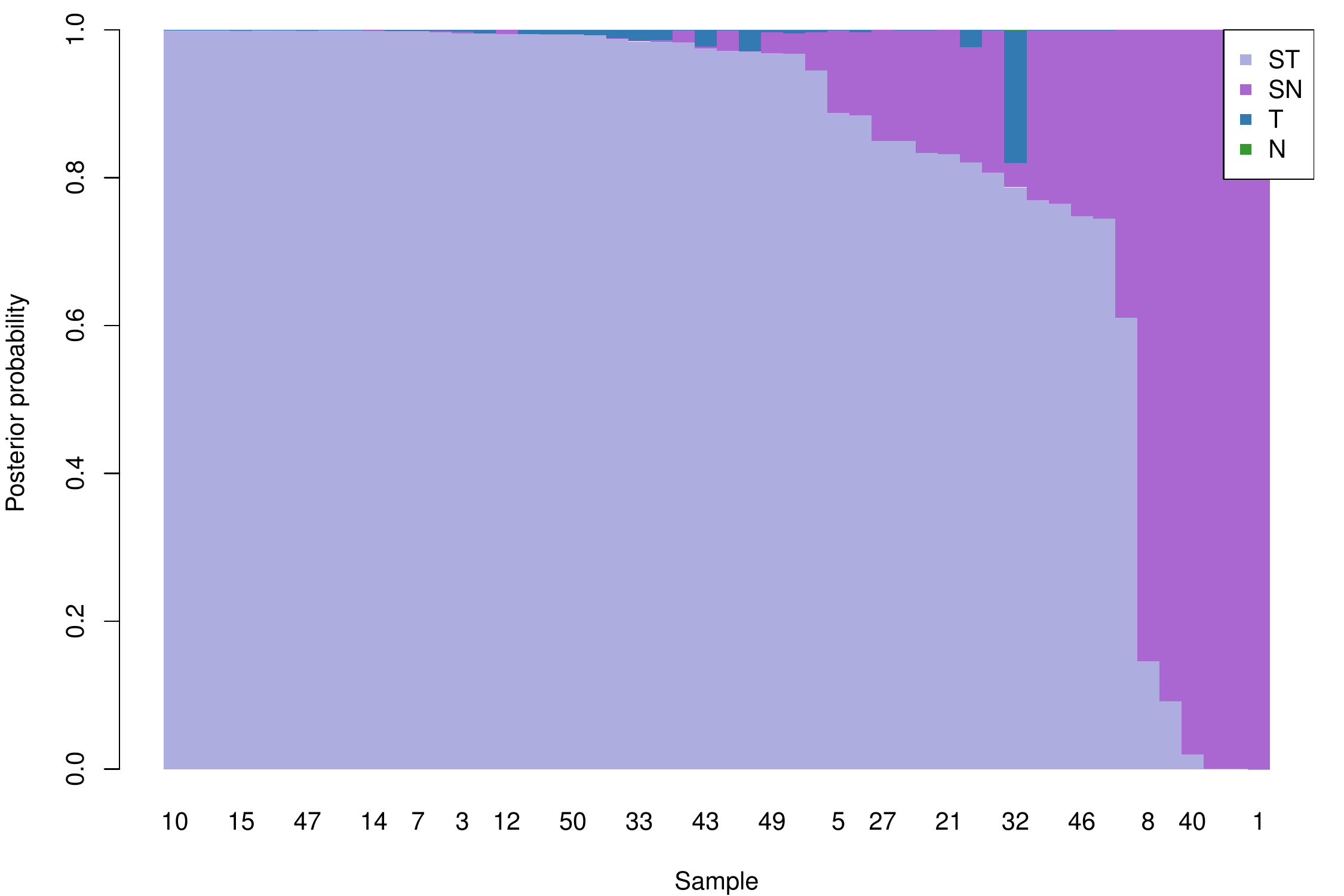}
 \caption{Simulation results for the \textit{ST}-distributed samples.}
 \label{fig:postProbsST}
\end{figure}
\newline The worst performance of our approach happens when the data are generated from a \textit{SN} distribution. In Fig.\,\ref{fig:postProbsSN}, it is possible to notice that the procedure detects the correct \textit{SN} model in about 25\% of the cases, and it more often prefers the multivariate normal model: this can be justified by the fact that the multivariate \textit{SN} model is notoriously the most difficult to deal with, because of the multimodality phe\-no\-me\-non, described in \cite{LisPar13}.\\

Also in the \textit{ST} case (Fig.\,\ref{fig:postProbsST}), the true model has been correctly identified in 44 cases. In almost all the other cases, the \textit{SN} model has been preferred.

\subsection{The \texttt{mvst} package}
The simulation results have been obtained in R, using the package \texttt{mvst}. It contains functions to estimate the parameters of the \textit{ST} (and nested) models, and to simulate data from them. It uses the model and the proposals described above, even if it allows to define customized prior and proposal distributions.\\
It makes use of the GNU Scientific Library (see \citealp{Go09}) to speed up the heaviest parts of the code and, in particular, for the computation of \eqref{eq:parabolic}. Besides, it requires three R packages: \texttt{mvtnorm} \citep{GeBr15}, \texttt{MCMCpack} \citep{MaQuPa11} and \texttt{mnormt} \citep{ag16}. It also makes use of three scripts available in the \texttt{RcppGSL} package \citep{EdRo15}.

\section{A real dataset}
As a final illustration of the proposed algorithm, we consider the wine data of the Grignolino cultivar, used in \S6.2.6 of \citet{acbook}. The dataset contains 71 observations on 3 variables (chloride, glycerol and magnesium). Data are available in the \texttt{sn} package.
\begin{table}[h]
\begin{center}
 \begin{tabular}{ccccc}
  \hline
  \rule{0cm}{0.4cm} Model & $N$ & Student-$t$ & \textit{SN} & \textit{ST}\\
  $\hat\pi(M|\bs y)$ & 6.60e-14 & 2.22e-01 & 3.87e-11 & 7.78e-01\\
  \hline
 \end{tabular}
\end{center}
 \caption{Models' posterior probabilities.}
 \label{estTab}
\end{table}
\newline We have performed a PMC sampler with 6 iterations, 20000 particles each. The posterior probabilities for the four models are given in Table (\ref{estTab}). Models with light tails have negligible probabilities, while the preferred model is \textit{skew-t}.\\
Given this model, the posterior mean for $\nu$ is approximately equal to 3.22, while the ML estimate in \citet{acbook} is equal to 3.4.

\appendix
\section{Proof of Proposition 2.0.1}
\label{proofa1}
(a): From one of the possible definitions of a multivariate \textit{ST} r.v., it is known that 
$\bs U\sim SN_p(\bs 0, \bs\alpha, \bs\Omega, \nu)$; since $\bs Y$ is a simple transformation of 
$\bs U$, its distribution is readily obtained.\\
(b): Start from $f(y,z,v) = f(v) f(z) f(y\mid z, v)$. By assumption, $f(z)$ is a standard Gaussian density, and
\begin{displaymath}
 \left(\bs Y \mid Z=z, V=v\right) = \left(\bs\xi + \bs\omega \bs U \mid Z=z, V=v\right) =
 \begin{cases}
  \bs\xi + \bs\omega \bs X v^{-1/2} & z \geq 0 \\
  \bs\xi - \bs\omega \bs X v^{-1/2} & z < 0 \\
 \end{cases}.
\end{displaymath}
Then, by using simple results on conditional Gaussian densities, one gets
\begin{displaymath}
 \left(\bs Y \mid Z=z, V=v\right) \sim \left\{
 \begin{array}{ll}
  N_p\left (\bs\xi + \bs\omega \bs\delta \displaystyle\frac{z}{\sqrt{v}}, \frac{1}{v}\bs\omega (\bs\Omega - \bs\delta\bs\delta^\prime) \bs\omega\right) & z \geq 0\\
  \rule{0cm}{0.75cm} N_p\left (\bs\xi - \bs\omega \bs\delta \displaystyle\frac{z}{\sqrt{v}}, \frac{1}{v}\bs\omega (\bs\Omega - \bs\delta\bs\delta^\prime) \bs\omega\right) & z < 0\\
 \end{array}
 \right.
\end{displaymath}
Hence the result in \eqref{strep}.

\section{Proposal distributions}\label{Appx:propDist}
We use the full conditional distributions as proposals for the latent variables $\bs Z$ and $\bs\xi$: each $Z_i$ has the following full conditional distribution
\begin{equation}
 \pi(z_i|\cdots) = \frac{\phi(z^+_i|m_i, v_\theta)}{2(1-\Phi(z_i|m_i,v_\theta))}
\end{equation}
where
\begin{eqnarray*}
  v_\theta & = & (1+\bs\psi'G^{-1}\bs\psi)^{-1}\\
  m_i & = & v_\theta\sqrt{v_i} (\bs\psi' G^{-1}(\bs y_i-\bs\xi))
\end{eqnarray*}
The variables $Z_i$ can be drawn as the product of $Z^+_i$, a normal r.v. with parameters $m_i$ and $v_\theta$ truncated in 0 and the sign $S_i$, uniform on $\{-1, 1\}$. To generate values $Z^+$ a rejection sampler has been employed (see \citealp{MR1963334}).\\

The parameter $\bs\xi$ has the following full conditional density:
\begin{displaymath}
 (\bs\xi|\cdots) \sim N_p\left(\displaystyle\frac{1}{\sum_{i=1}^n v_i} \left(\sum_{i=1}^n(v_i\bs y_i) - \bs\psi\sum_{i=1}^n \sqrt{v_i}|z_i|\right),\displaystyle\frac{1}{\sum_{i=1}^n v_i}\,G\right)
\end{displaymath}
The parameters $\bs\psi$ and $G$ have untractable full conditional distributions. To obtain a proposal distribution, they are approximated using only the contribution of the likelihood to the full conditional density.\\
The parameter $\bs\psi$ has the following full conditional distribution
\begin{eqnarray*}
 \pi(\bs\psi\mid\cdots) & \propto & \prod_{j=1}^p \left[(G_{jj} + \psi_j^2)^{-1/2}\right] \mathbbm 1_{\bs{\delta}}(\Delta_{\Sigma})\\
  & \times & \exp\Bigg\{-\displaystyle\frac 1 2 \sum_{i=1}^n v_i \left(\bs y_i-\bs\xi-\bs\psi\frac{|z_i|}{\sqrt{v_i}}\right)' G^{-1} \left(\bs y_i-\bs\xi-\bs\psi\frac{|z_i|}{\sqrt{v_i}}\right)\Bigg\},
\end{eqnarray*}
where $\mathbbm 1_{x}(\cdot)$ denotes the indicator function. By ignoring the first two factors, we obtain the following proposal distribution
\begin{displaymath}
 q(\bs\psi) = \phi_p \left(\bs\psi\, \Big| \frac{1}{\sum_{i=1}^n z_i^2}\sum_{i=1}^n |z_i|\sqrt{v_i}(\bs y_i-\bs\xi),\frac{1}{\sum_{i=1}^n z_i^2} G\right)
\end{displaymath}
The proposal distribution has a positive density on $\mathbb R^p$, while the full conditional is bounded on $\Delta_\Sigma$. This feature improves the ability of the sampler to explore the parameter space; moreover, particles which don't respect the constraint~(\ref{eq:constraint}) will be automatically discarded, as they have null prior (and posterior) probability density, hence a null importance weight.

The parameter $G$ has the following full conditional density
\begin{displaymath}
 \pi(G|\cdots) \propto \pi(\Sigma)|J| |G|^{-n/2}\exp\left\{-\frac 12 \tr(G^{-1}\Lambda)\right\}
\end{displaymath}
Ignoring the prior term we obtain
\begin{displaymath}
 q(G) = IW(n-p-1,\Lambda).
\end{displaymath}

\section{Details about the Rejection Sampler}\label{Appx:RS}
For a generic latent variable $V_i$, the Kullback Leibler divergence $KL(f||\pi_v)$ is given by
\begin{displaymath}
 KL(f||\pi_v) = \int_{\mathbb R^+} f(v_i)\log\left(\frac{k_v\beta_v^{\alpha_v}}{2\Gamma(\alpha_v)} v_i^{\alpha_v/2-C} \exp\{A_i v_i + (B_i-\beta_v)\sqrt{v_i}\}\right)dv_i
\end{displaymath}
which has an analytical solution for $\alpha_v^\star = 2C$:
\begin{displaymath}
 KL(f||\pi_v) = \log\left(\frac{k_v\beta_v^{2C}}{2\Gamma(2C)}\right) + \frac{2C(2C+1)A_i}{\beta_v^2} + 2C\log(\beta_v) + \frac{2CB_i}{\beta} - 2C.
\end{displaymath}
This divergence has always one (and only one) minimum in $\mathbb R^+$, given by
\begin{displaymath}
 \beta_v^\star = \frac{1}{2} \left(B_i + \sqrt{B_i^2 + 8 A_i (2C+1)}\right).
\end{displaymath}


\begin{thebibliography}{17}
\expandafter\ifx\csname natexlab\endcsname\relax\def\natexlab#1{#1}\fi

\bibitem[{Azzalini \& Capitanio (2014)}]{acbook}
\textsc{Azzalini, A. } (2014).
\newblock \textit{{The Skew-Normal and related families}},
\newblock (with the collaboration of A. Capitanio).
\newblock Cambridge: Cambridge University Press.

\bibitem[{Azzalini(2015)}]{Azza15}
\textsc{Azzalini, A.} (2015).
\newblock \textit{{The {R} package \texttt{sn}: The Skew-Normal and Skew-$t$
  distributions (version 1.3-0)}}.
\newblock Universit{\`a} di Padova, Italia.

\bibitem[{Azzalini \& Capitanio(2003)}]{ac03}
\textsc{Azzalini, A.} \& \textsc{Capitanio, A.} (2003).
\newblock {Distributions generated by perturbation of symmetry with emphasis on a multivariate skew \textit{t} distribution}.
\newblock \textit{Journal of the Royal Statistical Society, B}, 65, 367--389.

\bibitem[{Azzalini \& Genton(2008)}]{ag:08}
\textsc{Azzalini, A.} \& \textsc{Genton, M.} (2008).
\newblock {Robust likelihood methods based on the skew-t and related distributions}.
\newblock \textit{International Statistical Review}, 76, 106--119

\bibitem[{Azzalini \& Genz(2016)}]{ag16}
\textsc{Azzalini, A.} \& \textsc{Genz, A.} (2016).
\newblock {The {R} package \texttt{mnormt}: The multivariate normal
      and $t$ distributions (version 1.5-4)}.
\newblock \textit{http://azzalini.stat.unipd.it/SW/Pkg-mnormt}

\bibitem[{Branco \& Dey(2001)}]{dey:01}
\textsc{Branco, M. D.} \& \textsc{Dey, D.} (2001).
\newblock A general class of multivariate skew-elliptical distributions.
\newblock \textit{{Journal of Multivariate Analysis}}, 77, 1--15.

\bibitem[{Branco et~al.(2011)}]{brgeli}
\textsc{Branco, M.D.}, \textsc{Genton, M.G.} \& \textsc{Liseo, B.} (2011).
\newblock Objective Bayesian Analysis of Skew-t Distributions.
\newblock \textit{Scandinavian Journal of Statistics}, 40 (1), 63--85

\bibitem[{Capp{\'e} et~al.(2004)}]{MR2109057}
\textsc{Capp{\'e}, O.}, \textsc{Guillin, A.}, \textsc{Marin, J.~M.} \&
  \textsc{Robert, C.~P.} (2004).
\newblock {Population {M}onte {C}arlo}.
\newblock \textit{J. Comput. Graph. Statist.} \textbf{13}, 907--929.

\bibitem[{Eddelbuettel \& Romain(2015)P}]{EdRo15}
\textsc{Eddelbuettel, D.} \& \textsc{Romain, F.} (2013).
\newblock {RcppGSL: 'Rcpp' Integration for 'GNU GSL' Vectors and Matrices}.
\newblock http://CRAN.R-project.org/package=RcppGSL.

\bibitem[{Fernandez \& Steel(1999)}]{FerSte99}
\textsc{Fernandez, C.} \& \textsc{Steel, M. F.~J.} (1999).
\newblock {Multivariate {S}tudent-{$t$} regression models: pitfalls and
  inference}.
\newblock \textit{Biometrika} \textbf{86}, 153--167.

\bibitem[{Fonseca \& al.(2008)}]{fonseca}
\textsc{Fonseca, T. C.}, \textsc{Ferreira, M. A. R.} \& \textsc{Migon, H. S.} (2008). 
\newblock {Objective Bayesian analysis for the Student-t
regression model}.
\newblock \textit{Biometrika} \textbf{95}, 325--333.

\bibitem[{Genton(2004)}]{gentbook}
\textsc{Genton, M.G.} (2004).
\newblock \textit{Skew-Elliptical Distributions and Their Applications: A Journey Beyond Normality}.
\newblock Genton, M.G. (Ed.). CRC/Chapman \& Hall/CRC, Boca Raton, FL.

\bibitem[{Genz et~al.(2015)}]{GeBr15}
\textsc{Genz, A.}, \textsc{Bretz, F.}, \textsc{Miwa, T.}, \textsc{Mi, X.},
  \textsc{Leisch, F.}, \textsc{Scheipl, F.} \& \textsc{Hothorn, T.} (2015).
\newblock \textit{{{mvtnorm}: Multivariate Normal and t Distributions}}.
\newblock R package version 1.0-3.

\bibitem[{Gough(2009)}]{Go09}
\textsc{Gough, B.} (2009).
\newblock \textit{{GNU Scientific Library Reference Manual - Third Edition}}.
\newblock Network Theory Ltd., (3rd ed.)

\bibitem[{Gradshteyn \& Ryzhik(1994)}]{GrRy94}
\textsc{Gradshteyn, I.~S.} \& \textsc{Ryzhik, I.~M.} (1994).
\newblock \textit{{Table of integrals, series, and products}}.
\newblock Academic Press, Inc., Boston, MA, russian ed.
\newblock Translation edited and with a preface by Alan Jeffrey.

\bibitem[{Hansen(1994)}]{han94}
\textsc{Hansen, B.E.} (1994).
\newblock Autoregressive conditional density estimation.
\newblock \textit{{Intern. Econ. Rev.}}, 35, 3, 705--730.

\bibitem[{Jones \& Faddy(2003)}]{jf03}
\textsc{Jones, M.C.} \& \textsc{Faddy, M.J.} (2003)
\newblock A Skew Extension of the t-Distribution, with Applications.
\newblock \textit{Journal of the Royal Statistical Society, B}, 65, 1, 159--174

\bibitem[{Lee \& McLachlan(2013)}]{LeMc13}
\textsc{Lee, S.~X.} \& \textsc{McLachlan, G.~J.} (2013).
\newblock {{EMMIXuskew}: An {R} Package for Fitting Mixtures of Multivariate
  Skew $t$ Distributions via the {EM} Algorithm}.
\newblock \textit{Journal of Statistical Software} \textbf{55}, 1--22.

\bibitem[{Leisen et~al.(2016)}]{leisen-skew}
\textsc{Leisen, F.,  Marin, J.M.} \& \textsc{Villa, C.} (2016).
\newblock \textrm{Objective Bayesian modelling of insurance risks with the skewed
Student-t distribution}.
\newblock \textit{Manuscript under preparation}.

\bibitem[{Liseo \& Parisi(2013)}]{LisPar13}
\textsc{Liseo, B.} \& \textsc{Parisi, A.} (2013).
\newblock {Bayesian inference for the multivariate skew-normal model: a
  population {M}onte {C}arlo approach}.
\newblock \textit{Comput. Statist. Data Anal.} \textbf{63}, 125--138.

\bibitem[{Marchenko \& Genton(2010)}]{MarGen2010}
\textsc{Marchenko, Y.} \& \textsc{Genton, M.} (2010).
\newblock {A suite of commands for fitting the skew-normal and skew-t models}.
\newblock \textit{Stata Journal} \textbf{10}, 507--539.
\newblock Cited By 2.

\bibitem[{Martin et~al.(2011)}]{MaQuPa11}
\textsc{Martin, A.~D.}, \textsc{Quinn, K.~M.} \& \textsc{Park, J.~H.} (2011).
\newblock {{MCMCpack}: Markov Chain Monte Carlo in {R}}.
\newblock \textit{Journal of Statistical Software} \textbf{42}, 22.

\bibitem[{Prates et~al.(2013)}]{PrBa13}
\textsc{Prates, M.~O.}, \textsc{Cabral, C. R.~B.} \& \textsc{Lachos, V.~H.}
  (2013).
\newblock {{mixsmsn}: Fitting Finite Mixture of Scale Mixture of Skew-Normal
  Distributions}.
\newblock \textit{Journal of Statistical Software} \textbf{54}, 1--20.

\bibitem[{{R Core Team}(2015)}]{RCT15}
\textsc{{R Core Team}} (2015).
\newblock \textit{{R: A Language and Environment for Statistical Computing}}.
\newblock R Foundation for Statistical Computing, Vienna, Austria.

\bibitem[{Rachev et~al.(2008)}]{rachevbook}
\textsc{Rachev, S.T.}, \textsc{Hsu, J.S.J.}, \textsc{Bagasheva, B.S.} \& \textsc{Fabozzi, F.J.} (2008).
\newblock \textit{Bayesian Methods in Finance}.
\newblock Wiley, New York.

\bibitem[Robert(1995)]{MR1963334}
\textsc{Robert, C.~P.} (1995).
\newblock {Simulation of truncated normal variables}.
\newblock \textit{Statistics and Computing}, 5(2):121--125, June 1995.

\bibitem[{Robert \& Casella(2004)}]{RoCa04}
\textsc{Robert, C.~P.} \& \textsc{Casella, G.} (2004).
\newblock \textit{{Monte {C}arlo statistical methods}}.
\newblock {Springer Texts in Statistics}. Springer-Verlag, New York, 2nd ed.

\bibitem[{Rosco et~al.(2011)}]{rosco}
\textsc{Rosco, J.F.}, \textsc{Jones, M.C.} \& \textsc{Pewsey, A.} (2011).
\newblock Skew t distributions via the sinh-arcsinh transformation.
\newblock \textit{TEST}, 20, 3, 630--652.

\bibitem[{Rubio et~al.(2015)}]{rubio}
\textsc{Rubio, F.J. \& Steel, M.F.J.} (2015).
\newblock Bayesian modelling of skewness and kurtosis with Two-Piece Scale and shape distributions.
\newblock \textit{Electron. J. Statist.}, 9, 2, 1884--1912.

\bibitem[{Sartori(2006)}]{sartori}
\textsc{Sartori, N.} (2006). 
\newblock Bias prevention of maximum likelihood estimates for scalar skew normal and skew-t distributions. 
\newblock \textit{J. Statist. Plan. Inference} 136, 4259--4275.

\bibitem[{StataCorp.(2015)}]{stata14}
\textsc{StataCorp.} (2015).
\newblock {Stata Statistical Software: Release 14.}

\end{thebibliography}
\end{document}